\documentclass{elsarticle}

\usepackage{latexsym}
\usepackage{float}
\usepackage[latin2]{inputenc}

\usepackage[hidelinks]{hyperref}

\makeatletter
\def\ps@pprintTitle{%
	\let\@oddhead\@empty
	\let\@evenhead\@empty
	\def\@oddfoot{\centerline{\thepage}}%
	\let\@evenfoot\@oddfoot}
\makeatother
\begin{document} 

\bigskip

\title{SECLAF: A Webserver and Deep Neural Network Design Tool for Biological Sequence Classification}

\author[p]{Balázs Szalkai\corref{cor1}}
\ead{szalkai@pitgroup.org}
\author[p,u]{Vince Grolmusz\corref{cor1}}
\ead{grolmusz@pitgroup.org}
\cortext[cor1]{Corresponding authors}
\address[p]{PIT Bioinformatics Group, Eötvös University, H-1117 Budapest, Hungary}
\address[u]{Uratim Ltd., H-1118 Budapest, Hungary}

\date{}


\begin{abstract}
 Artificial intelligence (AI) tools are gaining more and more ground each year in bioinformatics. Learning algorithms can be taught easily by using the existing enormous biological databases, and the resulting models can be used for the high-quality classification of novel, un-categorized data in numerous areas, including biological sequence analysis.
Here we introduce SECLAF, an artificial neural-net based biological sequence classifier framework, which uses the Tensorflow library of Google, Inc. By applying SECLAF for residue-sequences, we have reported (Methods (2017), \url{https://doi.org/10.1016/j.ymeth.2017.06.034}) the most accurate multi-label protein classifier to date (UniProt --into 698 classes-- AUC 99.99\%; Gene Ontology --into 983 classes-- AUC 99.45\%). Our framework SECLAF can be applied for other sequence classification tasks, as we describe in the present contribution.

\noindent{\bf Availability and implementation:} The program SECLAF is implemented in Python, and is available for download, with example datasets at the website \url{https://pitgroup.org/seclaf/}. For Gene Ontology and UniProt based classifications a webserver is also available at the address above. 
\end{abstract}

\maketitle

\section{Introduction and motivation} 

New biological sequences are identified and submitted to public depositories by the thousands every day. The classification and annotation of these data is a demanding task. One possible solution for sequence classification could be the application of advanced artificial intelligence tools, such as artificial neural networks \citep{McCulloch1943}. In a previous work \citep{Seclaf} we have constructed a framework, called SECLAF (Sequence Classification Framework), and have demonstrated its considerable power by multi-label classification of UniProt and Gene Ontology entries. As we have demonstrated in \citep{Seclaf}, the SECLAF produces the most accurate artificial neural network for residue sequence classification to date (for UniProt --into 698 classes-- AUC 99.99\%; for Gene Ontology --into 983 classes-- AUC is 99.45\%).

Here we publish the downloadable SECLAF program, and, additionally, a pre-configured webserver at \url{https://pitgroup.org/seclaf/}.  Our goal was to create a tool for designing deep neural networks which classify biological sequences. We wanted to reduce the amount of programming for the users of SECLAF, which we achieved by requiring only that the input dataset (training and testing data) should be in a certain format, but the neural network architecture and  hyperparameters can be supplied in a human-readable JSON file. Preparation of the input data must be done by the user, but after that, no more coding is required.

\section{Materials and methods}

We implemented SECLAF in Python, using the neural network library Tensorflow \citep{Abadi2016a,Abadi2016c,Rampasek2016}. Python is a widely used language in bioinformatics research. Tensorflow is a relatively new framework created by Google which allows one to define and train neural networks at a low or higher level. We chose Tensorflow because it is easy to install, supports low level operations which eliminates the need for writing CUDA code when defining new layers, and performs automatic differentiation. In addition, it is sufficiently fast when compared to the other options.

\section{Implementation and usage}

When training a neural network, the input of SECLAF should consist of the following files:

\begin{itemize}
	\item the tree file: a hierarchy of the sequence classes (\texttt{classes.tre}),
	
	\item the training set: a file containing the training sequences along with their classification (\texttt{train\_set.ann}),
	
	\item the test set: a file with the testing sequences and their annotations (classifications), which must be a file with the same structure as \texttt{train\_set.ann},
	
	\item and a file containing the network configuration and various necessary parameters for training, testing and inference (\texttt{config.json}).
\end{itemize}

SECLAF can perform a hierarchical classification of sequences. This includes non-hierarchical classification as a special case: if the classes are pairwise disjoint and there is no implication between class membership, the class hierarchy file should only contain a list of all classes with no parent classes specified, along with the textual descriptions of the classes. However, if there is an implication relationship between some or all of the classes, e.g. they can be organized into a tree with superclasses and subclasses, then this file should also contain the logical implications between class memberships. For example, if all sequences in class A also belong to class B, then being a member of class A logically implies being a member of class B. In other words, a relation can be defined on the classes, which we will call \texttt{is\_a} after Gene Ontology terminology. Then \texttt{A is\_a B} would mean that all the sequences in class A are members of class B as well. In this case, the line describing class A in the tree file must also contain a list of all the classes that are implied by A, i.e. those classes X for which \texttt{A is\_a X}.

SECLAF will use the information about class hierarchy in both training and inference. In the training and test sets, the superclasses do not have to be present if the sequences are properly annotated with their corresponding subclasses because SECLAF will auto-complete the annotations by including all parent classes and their parents. In addition, when doing inference, SECLAF will output a subgraph of the class hierarchy with no outgoing edges, meaning that if class A is an output for some sequence S, and another class B is (indirectly or directly) implied by A, then B will also be present in the output corresponding to sequence S. The exact format required for the class hierarchy file (and also for the sequence container files) is described in the readme file of SECLAF.

In the configuration file, one can configure basic neural network hyperparameters such as the learning rate, the learning rate decay schedule, the multiplier for L2 regularization loss, the batch size, the number of iterations, and parameters concerning class balancing. Constraints on the input sequences and classes can also be given: their minimum length (the neural network will have a lower bound on sequence length depending on the architecture), maximum length (if overly long sequences cannot fit into GPU memory), the minimum class size (number of sequences), and the maximum depth in the class hierarchy to consider.

The input sequence encoding must also be specified in the configuration file, as the neural network cannot accept character sequences, only numeric values. SECLAF can encode both DNA and protein sequences, but they cannot be mixed. Multiple encoding methods are available. The most simple one (SimpleDnaEncoder, SimpleAminoAcidEncoder) assigns the elements of a 4- or 20-dimensional standard basis (i.e. one-hot vectors) to each nucleotide or amino acid. A compact encoding method is available for DNA sequences (CompactDnaEncoder), which assigns a 3-dimensional vector to each nucleotide: the three components are all binary and correspond to the purine/pyrimidine, strong/weak and amino/keto dichotomies. Another method (CompactAminoAcidEncoder) assigns a 6-dimensional vector to each amino acid based on its chemical properties, and the last one (BigAminoAcidEncoder) assigns the concatenation of the two kinds of vectors (20- and 6-dimensional) to each amino acid, thus yielding a 26-dimensional vector. For example, BigAminoAcidEncoder will assign a matrix of size $L \times 26$ to an input sequence with length $L$. If $N$ denotes the minibatch size, then a whole minibatch of sequences will be assigned a 3-rank array with shape $N \times L \times 26$.

The network architecture must also be configured in the \texttt{config.json} file. The architecture should be given as a list of layers, excluding the input layer. The last one in the list will be the output layer, which must be a fully connected layer with the same number of outputs as the number of classes selected for classification. SECLAF supports the following layers: 1-dimensional convolution, 1-dimensional max pooling, batch normalization, dropout, 1-dimensional spatial pyramid pooling, and fully connected (dense). As the input of the network has a variable length, while its output has a fixed length, there must be a spatial pyramid pooling layer at a point, after which only batch normalization, dropout, and fully connected layers are allowed. The spatial pyramid pooling (SPP) layer was introduced in \citep{Yang2009} for image processing. It seems to be a restriction that each network architecture given for SECLAF must contain an SPP layer, but in fact level-1 SPP corresponds to the simple special case when the activations are spatially max-pooled, resulting in a vector with as many dimensions as the number of filters in the last convolutional layer, which is a commonly used technique when using a neural network on variable-length sequences.

Two examples are available with SECLAF to demonstrate how to use the program. One example classifies proteins into 983 Gene Ontology classes; the other one is the same network architecture applied for protein classification into 698 UniProt families. These networks use all the supported layer types, so they provide a comprehensive example for describing a neural network in SECLAF.



\end{document}